\documentclass[useAMS,usenatbib,usegraphicx]{mn2e} 

\usepackage[usenames]{color}
\usepackage{amssymb}
\usepackage{amsmath}
\usepackage{bm}
\usepackage{comment}

\newcommand{\be}{\begin{equation}}
\newcommand{\ee}{\end{equation}}
\newcommand{\bea}{\begin{eqnarray}}
\newcommand{\eea}{\end{eqnarray}}
\newcommand{\lm}{\Lambda{\rm CDM}}

\newcommand{\hdist}{\,h^{-1}\,\mathrm{Mpc}}

\newcommand{\MR}{M|\lambda}

\newfont{\gwpfont}{cmssq8 scaled 1000}

\newcommand{\Pval}{6.48\pm3.00}
\newcommand{\cval}{1.65\pm0.95}
\newcommand{\alphaval}{0.87\pm0.25}
\newcommand{\betaval}{3.19\pm0.66}

\newcommand{\UPPbf}{[5.42,\,1.12,\,0.31,\,0.81,\,3.47]}

\newcommand{\AYM}{-0.22 \pm 0.04}
\newcommand{\BYM}{1.72 \pm 0.07}

\newcommand{\bias}{0.76 \pm 0.05}
\newcommand{\alphaMR}{1.22 \pm 0.04}
\newcommand{\logMRval}{14.432 \pm 0.041}

\newcommand{\logMMelchior}{14.327 \pm 0.044}
\newcommand{\logMSimet}{14.329 \pm 0.043}
\newcommand{\logMBaxter}{33.183 \pm 0.099}

\title[Cluster Scaling Relations]{\textit{Planck}/SDSS Cluster Mass and Gas Scaling Relations for a Volume-Complete redMaPPer Sample}

\author[Jimeno et al.]
{Pablo Jimeno$^{1}$\thanks{E-mail: pablodavid.jimeno@ehu.eus}, Jose M. Diego$^{2}$, Tom Broadhurst$^{1,3}$, I. De Martino$^{1}$, Ruth Lazkoz$^{1}$\\
$^{1}$Department of Theoretical Physics and History of Science, University of the Basque Country UPV-EHU, 48040 Bilbao, Spain\\
$^{2}$Instituto de F\'isica de Cantabria (CSIC-UC), Avda. Los Castros s/n, 39005 Santander, Spain\\
$^{3}$IKERBASQUE, Basque Foundation for Science, Alameda Urquijo, 36-5 48008 Bilbao, Spain\\}

\date{Draft version \today} 

\pagerange{\pageref{firstpage}--\pageref{lastpage}}

\begin{document}

\maketitle

\label{firstpage} 

%%%%%%%%%%%%%%%%%%%%%%%%%%%%%%%%%%%%%%%%%%%%%%%%%%%%%%%%%%%%%%%%%%%%%%%%%%%%%%%
\begin{abstract} 
Using {\it Planck} satellite data, we construct SZ gas pressure profiles for a large, volume-complete sample of optically selected clusters. We have defined a sample of over 8,000 redMaPPer clusters from the Sloan Digital Sky Survey (SDSS), within the volume-complete redshift region $0.100 < z < 0.325$, for which we construct Sunyaev-Zel'dovich (SZ) effect maps by stacking {\it Planck} data over the full range of richness. Dividing the sample into richness bins we simultaneously solve for the mean cluster mass in each bin together with the corresponding radial pressure profile parameters, employing an MCMC analysis. These profiles are well detected over a much wider range of cluster mass and radius than previous work, showing a clear trend towards larger break radius with increasing cluster mass. 

Our SZ-based masses fall $\sim$24\% below the mass--richness relations from weak lensing, in a similar fashion as the ``hydrostatic bias'' related with X-ray derived masses. We correct for this bias to derive an optimal mass--richness relation finding a slope $\alphaMR$ and a pivot mass $\log_{10}(M_{500}/M_{\odot})= \logMRval$, evaluated at a richness $\lambda=60$. Finally, we derive a tight $Y_{500}$--$M_{500}$ relation over a wide range of cluster mass, with a power law slope equal to $\BYM$, that agrees well with the independent slope obtained by the {\it Planck} team with an SZ-selected cluster sample, but extends to lower masses with higher precision.
\end{abstract}

%%%%%%%%%%%%%%%%%%%%%%%%%%%%%%%%%%%%%%%%%%%%%%%%%%%%%%%%%%%%%%%%%%%%%%%%%%%%%%%

%%%%%%%%%%%%%%%%%%%%%%%%%%%%%%%%%%%%%%%%%%%%%%%%%%%%%%%%%%%%%%%%%%%%%%%%%%%%%%%
\section{Introduction}
\label{sect:introduction}

Galaxy clusters are powerful cosmological probes that provide complementary constraints in the era of ``Precision Cosmology''. They contribute accurate consistency checks and unique new competitive constraints because of the well understood cosmological sensitivity of their numbers and clustering \citep{Jain2013, Huterer2015, Dodelson2014, Pouri2014, Pan2015}. The growth of structures has led the observational evidence to support dark energy dominance today, in combination with complementary constraints
\citep{Efstathiou1990, Lahav1991, Bahcall2000, Allen2011, Carroll1992, Ostriker1995,
Bahcall1998}. To realize their full cosmological potential, large, homogeneous samples of clusters are now being constructed out to $z\simeq 1$ with weak lensing based masses, in particular the Subaru/HSC and JPAS surveys \citep{Oguri2017, Benitez2014, Jimeno2015, Jimeno2017}.

Currently the best direct lensing masses are limited to relatively small subsamples of X-ray and Sunyaev-Zel'dovich (SZ) selected clusters, totalling $\sim$100 clusters \citep{Umetsu2014, Zitrin2015, Okabe2016}. One of the main efforts is focused on defining scaling relations between clusters with such weak lensing masses and the more widely available X-ray, SZ and/or optical richnesses with the reasonable expectation that these relations may provide mass proxies in the absence of direct lensing masses. Such proxies have a physical basis for clusters that appear to be virialised, so that X-ray temperature and emissivity profiles can provide virial masses under hydrostatic equilibrium. Independently, the SZ distortion of the cosmic microwave background (CMB) spectrum relates the density and temperature of cluster gas through inverse Compton scattering, and hence naturally anticipated to scale approximately with cluster mass. The cleanest mass proxy is arguably provided by the number of member galaxies, the so called richness, implicit in the assumption that the dominant cluster dark matter is collisionless like galaxies, and indeed the mass and the richness are found to be closely proportional \citep{Bahcall1977, Girardi2000}.

The cluster mass--richness relation, crucial in any attempt to use large number of clusters detected in the optical to constrain cosmological parameters, has been estimated in the past decade using cluster catalogues derived from the Sloan Digital Sky Survey (SDSS, \citealt{Gunn2006}) data, like the MaxBCG \citep{Koester2007}, the GMBCG \citep{Hao2010}, the WHL12 \citep{Wen2012}, or the more sophisticated redMaPPer cluster catalogue, both in its SDSS \citep{Rykoff2014} and DES \citep{Rykoff2016} versions. This relation can be estimated directly obtaining cluster masses from X-rays, weak lensing, SZ effect or velocity dispersion measurements in clusters \citep{Johnston2007, Andreon2010, Saro2015, Sereno2015a, Saro2016, Simet2016, Mantz2016, Melchior2016}, or indirectly, using numerical simulations \citep{Angulo2012, Campa2017} or comparing the observed abundances or clustering amplitudes with model predictions \citep{Rykoff2012, Baxter2016, Jimeno2017}.

The dynamical evolution and growth of galaxy clusters are driven by the dominant dark matter, but the relevant observables depend on the physical state of the baryons. Hence scaling relations between clusters observables and mass are not direct, but have been predicted to follow physically self-similar relations \citep{Kaiser1986, Kravtsov2012} that have been tested observationally and with hydrodynamical N-body simulations. Specifically, the integrated thermal SZ effect \citep{Sunyaev1972}, the X-ray luminosity, and the temperature are predicted to scale with the mass of the galaxy clusters as $Y\propto M^{5/3}$, $L_X\propto M^{4/3}$ and $T\propto M^{2/3}$, respectively. While simulations agree with the self-similar model \citep{White2002, daSilva2004, Motl2005, Nagai2006, Wik2008, Aghanim2009}, X-ray and SZ observations have uncovered departures from self-similarity that may be explained by complications due to cluster mergers, including shocked gas, cool gas cores, and energy injection from active galactic nuclei (AGN)\citep{Voit2005, Arnaud2005, Arnaud2007, Pratt2009, Vikhlinin2009, DeMartino2016}. Differences between the observations and purely gravitationally predicted scaling relations then provide insights into the interesting physics of the intracluster medium \citep{Bonamente2008, Marrone2009, Arnaud2010, Melin2011, Andersson2011, Comis2011, Czakon2015}.

In practice, samples of strong SZ-selected clusters that are also bright X-ray sources are currently being used to calibrate the SZ--mass relation \citep{Arnaud2010, Planck2013V, Planck2013XX, Saro2015}, but, since such clusters are often out of hydrostatic equilibrium for the reasons mentioned above, an SZ--mass scaling relation requires a correction for ``hydrostatic mass bias'' \citep{Nagai2007a, Zhang2010, Shi2014, Sayers2016}. This hot gas related bias can be broadened by other systematics like object selection process or by temperature inhomogeneities in X-ray measurements. Another approach to calibrate the masses of the cluster sample is to stack clusters in terms of richness and measure the SZ signal as a function of richness. This was done first by \cite{Planck2011XII} using the MaxBCG catalogue and stacking the \textit{Planck} data, and more recently by \cite{Saro2016} using an initial sample of 719 DES clusters with South Pole Telescope (SPT) SZ data and assuming various priors to extract the SZ signal.

In this work, we extract the \textit{Planck} SZ signal from $\sim$ 8,000 redMaPPer clusters identified in the SDSS, that have allowed us to previously define accurate clustering and density evolution measurements in the redshift range $0.100<z<0.325$, as described in detail in \cite{Jimeno2017}. Here we take this well defined cluster sample and stack the {\it Planck} multi-frequency data over a wide range of cluster richness. We only need to assume a weak prior for the global gas fraction using X-ray measurements to simultaneously derive a more ``self-sufficient'' method to derive SZ pressure profiles and the corresponding mean cluster masses binned by richness. Comparing masses derived this way with those expected from weak lensing mass--richness relations found in the literature, we derive the level of intrinsic bias for our sample and 
we then derive both a debiased mass--richness and a $Y_{500}$--$M_{500}$ relation describing our observational results.

This paper is organised as follows. In Sec.~\ref{sect:data} we describe the data that we use in our analysis, namely the redMaPPer cluster catalogue and {\it Planck} HFI maps. 
We present the basic theory associated to the SZ effect in Sec.~\ref{sect:sz_modeling}, as well as the mass--richness relation and the different models that we consider. In Sec.~\ref{sect:sz--mass_relation_analysis} we process the {\it Planck} data to obtain SZ maps given in terms of the Compton parameter $y$ and use them to constrain, through a joint likelihood analysis, the universal pressure profile parameters and the mean masses of the cluster subsamples considered. In Sec.~\ref{sect:mass--richness_relation_analysis} we make an estimation of the value of the mass bias and obtain the optimal mass--richness relation able to describe our bias-corrected masses. Finally, we use all the results obtained in the previous sections to derive a $Y_{500}$--$M_{500}$ relation in Sec.~\ref{sect:y500_m500_relation}, and present our conclusions in Sec.~\ref{sect:conclusions}.

Throughout this paper we adopt a fiducial flat $\lm$ cosmology with a matter density $\Omega_m=0.3$ and a Hubble parameter with a value today of $H_0 = 70\,{\rm km}\,{\rm s}^{-1}\,{\rm Mpc}^{-1}$. We also consider ${h \equiv H_0/(100\,{\rm km}\,{\rm s}^{-1}\,{\rm Mpc}^{-1})}$ and ${h_{70} \equiv H_0/(70\,{\rm km}\,{\rm s}^{-1}\,{\rm Mpc}^{-1})}$. Cluster masses are given in terms of an spherical overdensity $\Delta_{c(m)}$ with respect to the critical (mean) density of the Universe, $M_{\Delta{c(m)}}=(4/3)\pi \Delta \rho_{c(m)}(z)r_\Delta^3$. Unless stated otherwise, we refer to $M_{500 c}$ as $M_{500}$.

\section{Data}
\label{sect:data}

\subsection{redMaPPer cluster catalogue}

The pioneering Sloan Digital Sky Survey (SDSS, \citealt{Gunn2006}) combines photometric and spectroscopic observations and has mapped the largest volume of the Universe in the optical to date, covering around 14,000 deg$^2$ of the sky. The information obtained has been made public periodically via Data Releases (DR), and many different cluster catalogues have been carefully constructed using very different cluster-finder algorithms \citep{Koester2007, Hao2010, Wen2012}. We focus our analysis on one of these SDSS-based catalogues, the red-sequence Matched-filter Probabilistic Percolation cluster catalogue (redMaPPer, \citealp{Rykoff2014}), based on SDSS DR8 photometric data, as given in the public 6.3 version \citep{Rykoff2016}. We use this catalogue because it offers a very low rate of projection effects ($<5\%$, according to the thorough analysis of \cite{Rykoff2014}). We have previously compared this catalogue with other cluster catalogues and found the redMaPPer 
catalogue to provide highly consistent results when comparing gravitational redshift and the level of magnification bias \citep{Jimeno2015} 
and more recently we have also constructed a spectroscopically sample of redMaPPer clusters to construct the most precise cluster correlation functions and mass distributions to date \citep{Jimeno2017}, finding a high degree of consistency in relation to $\Lambda$CDM based predictions from the forefront MXXL simulations \citep{Angulo2012}.

The redMaPPer cluster finder algorithm, based on \cite{Rozo2009b} and \cite{Rykoff2012}, is able to find potential clusters within the raw photometric data, and provides an estimation of the sky position of the five most probable central galaxies (CGs), the redshift $z_\lambda$, and the richness $\lambda$ of each cluster. In this catalogue, the richness estimations relies on a multi-colour self-training procedure that calibrates the red-sequence as a function of redshift. The richness of a cluster is defined as:
\begin{equation}
\lambda=\sum p_{\rm mem}\,\theta_L\,\theta_R\,,
\end{equation}
where $p_{\rm mem}$ is the membership probability of each galaxy found near the cluster, and $\theta_L$ and $\theta_R$ are luminosity and radius-dependent weights. A more in-depth explanation of the algorithm features can be found in \cite{Rykoff2014} and \cite{Rozo2015b}.

The resulting redMaPPer catalogue covers an effective area of 10,401 deg$^2$, and contains 26,111 clusters in the $0.08 \leqslant z_{\rm photo} \leqslant 0.55$ redshift range. Finally, it should be mentioned that this catalogue is volume-complete up to $z \lesssim 0.33$, and has a richness cutoff of $\lambda/S(z)>20$, where $S(z)$ is the ``scale factor'' that relates the richness with the number of observed galaxies above the magnitude limit of the survey, $\lambda/S(z)$. This detection threshold corresponds to a mass limit of approximately $M_{500} \geqslant 0.7 \times 10^{14}\,M_\odot$.

\subsection{Planck SZ data}

We combine the above optically selected cluster sample from SDSS with the all-sky temperature maps derived by the {\it Planck} space mission \citep{Planck2013I}. Although these maps have already been used to construct catalogues of SZ sources \citep{Planck2013XXIX} and an all-sky Compton $y$ parameter map \citep{Planck2013XXI}, we reprocess them for our own purposes.

To obtain the Compton parameter maps required in our analysis, we use the {\it Planck} full mission High-Frequency Instrument maps (HFI, \citealt{Planck2013IX}) at 100, 143, 217, and 353 GHz. These maps are provided in {\gwpfont HEALPix} format \citep{Gorski2005}, with a pixelisation of $N_{\rm side}=$ 2,048, which correspond to a pixel resolution of $\sim1.7$ $\mathrm{arcmin}$. The {\it Planck} effective beams for each of the 100, 143, 217, and 353 GHz channels can be approximated by circular Gaussians with FWHM values of 9.66, 7.27, 5.01 and 4.86 $\mathrm{arcmin}$, respectively. To compute the contribution of the SZ signal in the {\it Planck} temperature maps, we also make use of the spectral transmission information of each of these frequency channels, as given in \cite{Planck2013IX}.

\section{Model}
\label{sect:sz_modeling}

\subsection{Thermal Sunyaev-Zel'dovich effect}

Here we briefly introduce the equations that describe the thermal Sunyaev-Zel'dovich (SZ) effect. For a derivation, we refer the reader to the papers of \cite{Sunyaev1980, Rephaeli1995}, or the more recent work by \cite{Birkinshaw1999}, \cite{Carlstrom2002} and \cite{Diego2002}. Ignoring relativistic corrections, the SZ spectral distortion of the CMB, expressed as temperature change, is:
\begin{equation}\label{eq:DeltaT_to_y}
\frac{\Delta T_{\rm SZ}}{T_{\rm CMB}} = g(x)\,y\,,
\end{equation}
where $x=(h\,\nu)/(k_{B}\,T_{\rm CMB})$ is the dimensionless frequency, $y$ is the Compton parameter, and
\begin{equation}\label{eq:g_factor}
g(x) = \left( x\,\frac{e^x+1}{e^x-1}-4\right)\,.
\end{equation}

The Compton parameter $y$ is equal to the optical depth, $\tau_e$, times the fractional energy gain per scattering, and is given by:
\begin{equation}
y=\frac{\sigma_T}{m_e\,c^2}\int_0^\infty P(l)\,{\rm d}l\,,
\end{equation}
where $\sigma_T$ is the Thomson cross section and $P$ is the intracluster pressure produced by free electrons. Integrating over the solid angle of the cluster one obtains the integrated Compton parameter:
\begin{equation}\label{eq:Y_integrated}
Y=\int_\Omega y\,{\rm d}\Omega = D_A^{-2}\,\frac{\sigma_T}{m_e\,c^2}\int_0^\infty {\rm d}l \int_{A_{\rm clu}} P(l)\,{\rm d}A\,,
\end{equation}
where $A_{\rm clu}$ is the area of the cluster in the plane of the sky, and $D_A(z)$ is the angular diameter distance at redshift $z$.

Assuming an spherical model for the cluster, we have that the Compton parameter $y$ at a distance $r$ from the center of the cluster is equal to:
\begin{equation}
y(r)=\frac{\sigma_T}{m_e\,c^2}\int_{-\infty}^\infty P\left(\sqrt{{r'}^2+r^2}\right)\,{\rm d}{r'}\,,
\end{equation}
and thus the integrated Compton parameter $Y$, obtained integrating to a distance $R$ from the center of the cluster, is given by:
\begin{equation}
Y(R)=\int_0^R 2\pi\,y(r)\,r\,{\rm d}r\,,
\end{equation}
which has units of $\mathrm{Mpc}^2$. It should be noted that, as $y$ is a projected along the line of sight quantity, $Y$ is the so called ``cylindrical'' integrated Compton parameter $Y^{\rm cyl}$, and not the ``spherical'' integrated Compton parameter, which would be obtained directly from the pressure profile doing:
\begin{equation}
Y^{\rm sph}(R)=\frac{\sigma_T}{m_e\,c^2}\int_0^R 4\pi\,P(r)\,r^2\,{\rm d}r\,.
\end{equation}
In practice, we work with $Y^{\rm cyl}$ when dealing with observations, as this is the quantity that can be measured from the data, and we use $Y^{\rm sph}$ when dealing with models. Once a pressure profile has been adopted, any measurement of $Y^{\rm cyl}(n\,r_{500})$ can be straightforwardly converted in terms of $Y^{\rm cyl}(r_{500})$, and the latter to $Y^{\rm sph}(r_{500})$. We refer the reader to appendix A of \cite{Melin2011} for a more detailed explanation of how to convert between definitions.

Finally, as $y$ is dimensionless, $Y$ can also be expressed in units of $\mathrm{arcmin}^2$:
\begin{equation}
Y\,[{\rm arcmin}^2]=D_A(z)^{-2}\,\left(\frac{60\times180}{\pi}\right)^2\,Y\,[{\rm Mpc}^2]
\end{equation}

From now on, we refer to $Y^{\rm sph}(r_{500})$ as $Y_{500}$, given in $\mathrm {Mpc}^2$ units.

\subsection{Pressure profile}
\label{sect:pressure_profile}

In this work we adopt the generalised NFW (GNFW) ``universal pressure profile'' proposed by \cite{Nagai2007a}, that has a flexible double power-law form:
\begin{equation}\label{eq:normal_pressure}
\mathbb{P}(x)=\frac{P_0}{(c_{500}\,x)^\gamma\,\left[1+(c_{500}\,x)^\alpha\right]^{(\beta-\gamma)/\alpha}}\,,
\end{equation}
where $x=r/r_{500}$ is the scaled dimensionless physical radius. The physical pressure is given by:
\begin{equation}\label{eq:pressure_profile}
P(x) = P_{500}\,\left(\frac{M_{500}}{3\times10^{14}\,h_{70}^{-1}\,M_\odot}\right)^{\alpha_p}\,\mathbb{P}(x)\,,
\end{equation}
where:
\begin{equation}
\begin{aligned}
P_{500} = &1.65 \times 10^{-3}\,E(z)^{8/3}\\
&\times \left(\frac{M_{500}}{3\times10^{14}\,h_{70}^{-1}\,M_\odot}\right)^{2/3}\,{\rm h_{70}^2}\,{\rm keV}\,{\rm cm}^{-3}\,,
\end{aligned}
\end{equation}
and ${\alpha_p=0.12}$ accounts for the deviation from the self-similar scaling model \citep{Arnaud2010}. The value of $r_{500}$ is given by:
\begin{equation}\label{eq:mass_radius_relation}
\frac{4}{3}\,\pi\,500\,\rho_c(z)\,r_{500}^3 = M_{500}\,,
\end{equation}
where $\rho_c(z)$ is the critical density of the Universe at redshift $z$, defined as:
\begin{equation}
\rho_c(z) = \frac{3H_0^2\,E(z)^2}{8\pi G}\,,
\end{equation}
and $E(z)^2 = \Omega_{m}(1+z)^3 + (1-\Omega_{m})$.

From Eq.~\ref{eq:normal_pressure}, it is clear that the slopes of the pressure profile are given, at different $r_{500}$-scaled distances, by $\gamma$ for $x\ll1/c_{500}$, $\alpha$ for $x\sim1/c_{500}$, and $\beta$ for $x\gg1/c_{500}$. In our analysis and following the approach by \cite{Planck2013V}, we leave $P_0$, $c_{500}$, $\alpha$, and $\beta$ as free parameters. The low resolution of the {\it Planck} data does not have the power to constrain $\gamma$, so we fix it to $\gamma=0.31$, value obtained by \cite{Arnaud2010} from a sample of 33 {\it XMM-Newton} X-ray local clusters in the $r<r_{500}$ range.

\subsection{Gas fraction}
\label{sect:gas_fraction}

To improve our analysis, we use established results regarding the global gas fraction $f_\mathrm{gas}$ in clusters, particularly, those by \cite{Pratt2009}, who derived a mass--gas fraction relation using precise hydrostatic mass measurements of 41 {\it Chandra} and {\it XMM-Newton} clusters \citep{Vikhlinin2006, Arnaud2007, Sun2009}, which is also in good agreement with the results obtained from the {\gwpfont REXCESS} sample \citep{Böhringer2007}. According to their analysis, these clusters, whose masses range from $10^{13}\,M_\odot$ to $10^{15}\,M_\odot$, follow the mean mass--gas fraction relation:
\begin{equation}\label{eq:pratt_gas_fraction}
\begin{aligned}
\ln \left(f_{{\rm gas},500}\,E(z)^{-3/2}\right) = & \left(-2.37 \pm 0.03\right) \\
& + \left(0.21\pm0.03\right) \ln \left(\frac{M_{500}}{2\times10^{14}\,M_\odot}\right)\,.
\end{aligned}
\end{equation}
To compute the gas fraction we first need to compute the gas mass:
\begin{equation}\label{eq:gas_fraction_1}
 M_{{\rm gas},500}=\int^{r_{500}}_0 \mu_e\,m_u\,n_e(r)\,4\pi\,r^2\,{\rm d}r\,,
\end{equation}
where $\mu_e=1.15$ is the mean molecular weight per free electron, $m_u$ is the atomic mass unit, and $n_e(r)$ is the electron density. Because the intra-cluster pressure is given by $P(r)=n_e(r)\,k_{B}\,T$, assuming an isothermal model for the cluster one can directly derive $M_{{\rm gas},500}$ from the adopted pressure profile (Eq.~\ref{eq:pressure_profile}). 

For the temperature, we use the mean mass--temperature relation given by \cite{Lieu2016}:
\begin{equation}\label{eq:MT_relation}
\begin{aligned}
\log_{10}\left(\frac{M_{500}\,E(z)}{h_{70}^{-1}\,M_\odot}\right) = & \left(13.57^{+0.09}_{-0.09}\right) \\
& + \left(1.67^{+0.14}_{-0.10}\right)\log_{10}\left(\frac{k_BT}{\rm keV}\right)\,,
\end{aligned}
\end{equation}
which was obtained combining weak lensing mass estimates with {\it Chandra} and {\it XMM-Newton} temperature data of 38 clusters from the XXL survey \citep{Pacaud2016}, 10 clusters from the COSMOS survey \citep{Kettula2013}, and 48 from the Canadian Cluster Comparison Project (CCCP, \citealt{Mahdavi2013, Hoekstra2015}), spanning a temperature range $T\simeq 1-10$ keV.

It is worth mentioning that if an isothermal model is assumed and we consider that $M_{\rm gas}\propto f_b\,M$, where $M$ is the total cluster mass and $f_b$ is the baryon gas fraction, from Eq.~\ref{eq:Y_integrated} we have that the integrated Compton parameter scales as $Y \propto f_b\,M\,T\, D_A^{-2}$. However, even clusters in hydrostatic equilibrium are not strictly isothermal, and temperatures are commonly observed to drop by a factor of $\sim$2 below a radius of $r \lesssim 100 - 200$ kpc because of strong radiative cooling, described best by a broken power law with a transition region \citep{Vikhlinin2006}. In any case, these scales are not resolved by {\it Planck} and in our analysis the assumption that the temperature is constant is a good approximation for the radial scales considered in this work.

\subsection{Miscentering}
\label{sect:miscentering}

In the redMaPPer catalogue, for each cluster the 5 most probable central galaxies (CGs) are provided with their corresponding centering probabilities. Usually, there is one CG with a much higher probability of being the real CG than the other 4, so we consider the most probable CG to be the center of the cluster. In any case, it is now known that, because clusters are still evolving systems, CGs do not always reside at the deepest part of the DM halo potential well \citep{vonderLinden2007}, but sometimes have high peculiar velocities, are displaced with respect to the peak of the X-ray emission \citep{Rozo2014a}, or are wrongly identified satellite galaxies \citep{Skibba2011}.

In stacked measurements on clusters, miscentering is one of the main sources of noise, and should be taken into account. When modelling the SZ signal coming from stacked samples of clusters, we introduce this effect considering the results obtained by \cite{Johnston2007}, who found a CG-center offset distribution that could be fitted by a 2D Gaussian with a standard deviation of $\sigma=0.42\hdist$ for the CGs that were not accurately centered, that occurs between 20 and 40 per cent of the time as a function of cluster richness, with a probability $p_{mc}(\lambda) = (2.13+0.046\,\lambda)^{-1}$.

However, it should be noted that this value of $0.42\,h^{-1}\,\mathrm{Mpc}$ is about 2 $\mathrm{arcmin}$ at $z \sim 0.2$, scale well below the resolution of the \textit{Planck} data we work with, so we do not expect this miscentering to introduce a high level of noise in our stacked measurements of the SZ effect.

\subsection{Mass bias}
\label{sect:HEbias}

Usually referred to as hydrostatic equilibrium (HE) masses, in their derivation there is an implicit assumption that the pressure is purely thermal. However, we may expect a non-negligible contribution to the total pressure from bulk and turbulent gas motions related to structure formation history, magnetic fields, and AGN feedback \citep{Shi2014, Planck2013XX}. Such non-thermal contributions to the total pressure would therefore cause masses estimated using X-ray or SZ observations to be biased low with uncertain estimates ranging between 5\% to 20\% \citep{Nelson2014, Nagai2007b, Rasia2006, Sembolini2013}.

We simply relate the HE mass estimates $M_{{\rm HE,}500}$ obtained from our SZ observations to true masses $M_{500}$ through a simple mass independent bias: 
\begin{equation}\label{eq:bias_factor}
 M_{{\rm HE,}500} = (1-b)\,M_{500}\,,
\end{equation}
where $(1-b)$ is the bias factor. This term can include not only the bias coming from departures from HE, but from observational systematics or sample selection effects.

\subsection{Mass--richness relation}
\label{sect:mass--richness_relation_model}

In order to explore the connection between the mass and the optical richness $\lambda$ in clusters, i.e., the number of galaxies contained within them, one needs to assume a form to describe the relation between cluster richness and mass. We consider the standard power law cluster mass--richness mean relation:
\begin{equation}\label{eq:MR_relation}
\left< M|\lambda \right> = M_0 \,\left(\frac{\lambda}{\lambda_0}\right)^{\alpha_{\MR}}\,,
\end{equation}
where $M_0$ is a reference mass at a given value of $\lambda=\lambda_0$, and $\alpha_{M|\lambda}$ is the slope of the mass--richness relation. In our case, we consider $\lambda_0=60$.

To compute the mean masses of our cluster subsamples, we first consider the probability $P(M|\lambda^{\rm obs})$ of having a given value of the mass $M$ for a cluster with $\lambda^{\rm obs}$:
\begin{equation}\label{eq:MR_relation_2}
P(M|\lambda^{\rm obs}) = \int P(M|\lambda)\,P(\lambda|\lambda^{\rm obs})\,{\rm d}\lambda\,,
\end{equation}
with $P(M|\lambda)$ a delta function, as the relation between mass and richness is given by Eq.~\ref{eq:MR_relation}. Following the usual approach \citep{Lima2005}, we consider that $P(\lambda|\lambda^{\rm obs})$ follows a log-normal distribution:
\begin{equation}\label{eq:MR_relation_3}
P(\lambda|\lambda^{\rm obs}) = \frac{1}{\sqrt{2\pi\sigma^2_{\ln \lambda|\lambda^{\rm obs}}}}\exp [-x^2(\lambda)]\,,
\end{equation}
where:
\begin{equation}
x(\lambda) = \frac{\ln \lambda - \ln \lambda^{\rm obs}}{\sqrt{2\sigma^2_{\ln \lambda|\lambda^{\rm obs}}}}\,,
\end{equation}
and $\sigma_{\ln \lambda|\lambda^{\rm obs}}$ is the fractional scatter on the halo richness at fixed observed richness, which is assumed to be constant with cluster redshift and richness. Because $P(M|\lambda)$ is a delta function, and considering Eq.~\ref{eq:MR_relation}, we also have that $\sigma^2_{\ln \lambda|\lambda^{\rm obs}} = \sigma^2_{\ln M|\lambda^{\rm obs}}$. So, the mean mass $\left< M \right>$ of one of the richness bins considered, with $\lambda \in [\lambda^{\rm obs}_i,\,\lambda^{\rm obs}_{i+1}]$ and containing $N_i$ clusters, is given by:
\begin{equation}
\left< M \right> = \frac{1}{N_i} \sum_{j=1}^{N_i} \int M\,P(M|\lambda^{\rm obs}_j)\,{\rm d} M\,.
\end{equation}

Finally, when dealing with mean values we can work in terms of $\left<\ln M|\lambda\right>$ instead of $\left< M|\lambda\right>$, with $\left<\ln M|\lambda\right> = \ln\left<M|\lambda\right> - 0.5\,\sigma^2_{\ln M|\lambda^{\rm obs}}$. We refer the reader to \cite{Rozo2009a} and \cite{Simet2016} for a discussion of this transformation. We choose our parametrisation (Eq.~\ref{eq:MR_relation}) because the resulting mean relation is less affected by the uncertainty in $\sigma_{\ln M|\lambda^{\rm obs}}$.

Since it was made public, there have been multiple attempts to constrain in different ways the parameters of this relation using the redMaPPer cluster catalogue \citep{Rykoff2012, Baxter2016, Farahi2016, Li2016, Melchior2016, Miyatake2016, Saro2015, Saro2016, Simet2016, Jimeno2017}. Although some of these works introduced a redshift dependence in the mass--richness relation, it was weakly constrained in all cases, and compatible with no redshift evolution at all. Given the small redshift range in which we work, redshift evolution is not important for our analysis and we refer our result to the mean redshift of our sample, $z=0.245$.

\section{Pressure profiles and mass estimation}
\label{sect:sz--mass_relation_analysis}

\subsection{Planck data processing}
\label{sect:planck_data_processing}

We first divide the redMaPPer cluster catalogue in 6 independent log-spaced richness bins, and take all those clusters that reside within the $0.100 < z < 0.325$ volume-complete redshift region. This leaves a total of 8,030 clusters, distributed in number and mean richness as shown in Table~\ref{tab:likelihood_masses}. 

Then, for each cluster subsample, we produce and stack the $\nu=$ 100, 143, 217 and 353 GHz 2.5~deg~$\times$~2.5~deg {\it Planck} maps associated to the clusters in each subsample and produce the corresponding SZ maps following a technique similar to the one used in \cite{Planck2016XL}, based on internal linear combinations (ILC) of the four different HFI maps. In our case we do not use the 70 GHz SZ map, as we prefer to smooth all the maps to a common higher 10 $\mathrm{arcmin}$ resolution instead. We also use the $M_{353}-M_{143}$ combination to clean the $M_{217}$ map, where $M_\nu$ is the {\it Planck} map at frequency $\nu$. To convert from $\Delta T_{\rm SZ}$ to $y$ units (Eq.~\ref{eq:DeltaT_to_y}), we compute the different effective spectral responses integrating the expected SZ spectrum (Eq.~\ref{eq:g_factor}) over each {\it Planck} bandpass.

\begin{figure}
\resizebox{84mm}{!}{\includegraphics{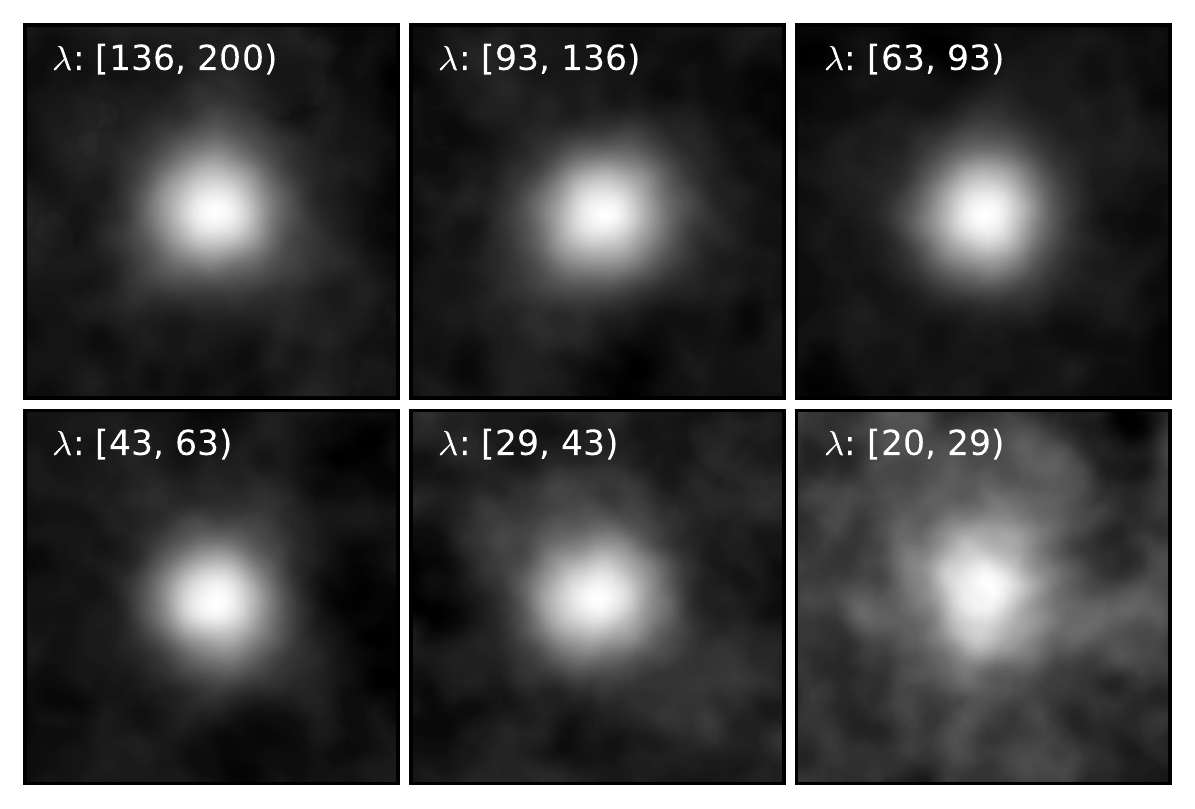}}
\caption{0.75~deg~$\times$~0.75~deg stacked Compton parameter $y$ maps in 6 richness bins (with the richness ranges indicated on each stamp), obtained through the process described in Sec.~\ref{sect:planck_data_processing}, showing the SZ effect is strongly detected in these cluster stacks over the full richness range. These subsamples are defined in the volume-complete redshift region $0.100 < z < 0.325$. The maps have been smoothed to a common 10 $\mathrm{arcmin}$ resolution.}
\label{fig:y_maps}
\end{figure}

Our final SZ maps of the cluster subsamples considered, given in terms of the Compton parameter $y$ and shown in Fig.~\ref{fig:y_maps}, are produced as a combination of the SZ$_{100}$ and SZ$_{143}$ maps, weighting them by the inverse of the variance of each map. This particular combination has been proposed by the \textit{Planck} team to maximise the signal-to-noise of the SZ component whilst minimising the contamination from Galactic emission and extragalactic infrared emission within clusters \citep{Planck2016XL}.

%================================================
%================================================

\subsection{Likelihood analysis}
\label{sect:likelihood_analysis_1}

We now combine the data from {\it Planck} and the constraints imposed by the gas fractions to perform a likelihood analysis that enables us to constrain the pressure profile parameters and the mean masses of the 6 cluster subsamples. We explore the values of our 4+6 dimensional model $\bm{\phi}=(P_0,\,c_{500},\,\alpha\,,\beta,\,M_{500}^1,\,...\,,\,M_{500}^{6})$ through a Monte Carlo Markov Chain (MCMC) analysis.

For a given value of $r_{500} (M_{500},\,z)$ we measure the Compton parameter $y$ within a disk of radius $x=0.35$ and in 6 annulus given by radii $x_i$ and $x_{i+1}$, where the $x_i=r_i/r_{500}$ values are log-spaced between 0.35 and 3.5. This results in a $y$ vector of 7 values. Then, to account for the background we subtract the mean value of the signal obtained from an annulus of radii $x_{\rm out}$ and $x_{\rm out} + x_{\rm FWHM}$, where $x_{\rm out}=3.5$, and $x_{\rm FWHM} = \theta_{\rm FWHM}\,D_A(z)/r_{500}$ corresponds in $x$-space to the $\theta_{\rm FWHM}=10$ $\mathrm{arcmin}$ FWHM effective resolution of the SZ maps. The values of $r_{500}$ used to both model the signal and measure it from the data are obtained from $M_{500}$ through Eq.~\ref{eq:mass_radius_relation}.

The log-likelihood employed has the form:
\begin{equation}
\ln\mathcal{L}(\bm{\phi}) = \sum_{k=1}^{6} \ln\mathcal{L}^k(\bm{y}^k|\bm{\phi}^k)\,,
\end{equation}
where $\bm{\phi}^k=(P_0,\,c_{500},\,\alpha\,,\beta,\,M_{500}^k)$, $\bm{y}^k$ is the data vector obtained from the cluster subsample $k$, and
\begin{equation}
\ln\mathcal{L}^k(\bm{y}^k|\bm{\phi}^k)\propto -\frac{1}{2}\,\chi^2(\bm{y}^k,\,\bm{\phi}^k,\,\mathcal{C})\,,
\end{equation}
where $\mathcal{C}$ is the covariance matrix, and
\begin{equation}\label{eq:chi_square}
\chi^2(\bm{y}^k,\,\bm{\phi}^k,\,\mathcal{C}) = \left(\bm{y}^k-\bm{\mu}(\bm{\phi}^k)\right) \mathcal{C}^{-1} \left(\bm{y}^k-\bm{\mu}(\bm{\phi}^k)\right)^T\,,
\end{equation}
with $\bm{\mu}$ the model values drawn from $\bm{\phi}^k$. To model the signal, for each $\bm{\phi}^k$ configuration we produce mock maps of the Compton parameter $y$ as a function of redshift. Then, we mimic miscentering effects adding to the mock maps the same maps smoothed with a 2D Gaussian of width $\sigma_{mc}=0.42\hdist$, and weighted by $p_{mc}(\lambda) = (2.13+0.046\,\lambda)^{-1}$. Finally, we produce a weighted map integrating over the redshift distribution of the subsample considered, convolve it with a 10 $\mathrm{arcmin}$ FWHM Gaussian, and perform the same measurements made in the {\it Planck} data maps.

It should be noted that the mock $y$ maps that we create to fit the observed signal are generated from the pressure profile as given by Eq.~\ref{eq:pressure_profile} and for a given total model mass $M^k_{500}$. Hence, to compute $\ln\mathcal{L}^k(\bm{y}^k|\bm{\phi}^k)$ we do not rely on the gas mass or $f_{gas}$.

\begin{figure}
\resizebox{84mm}{!}{\includegraphics{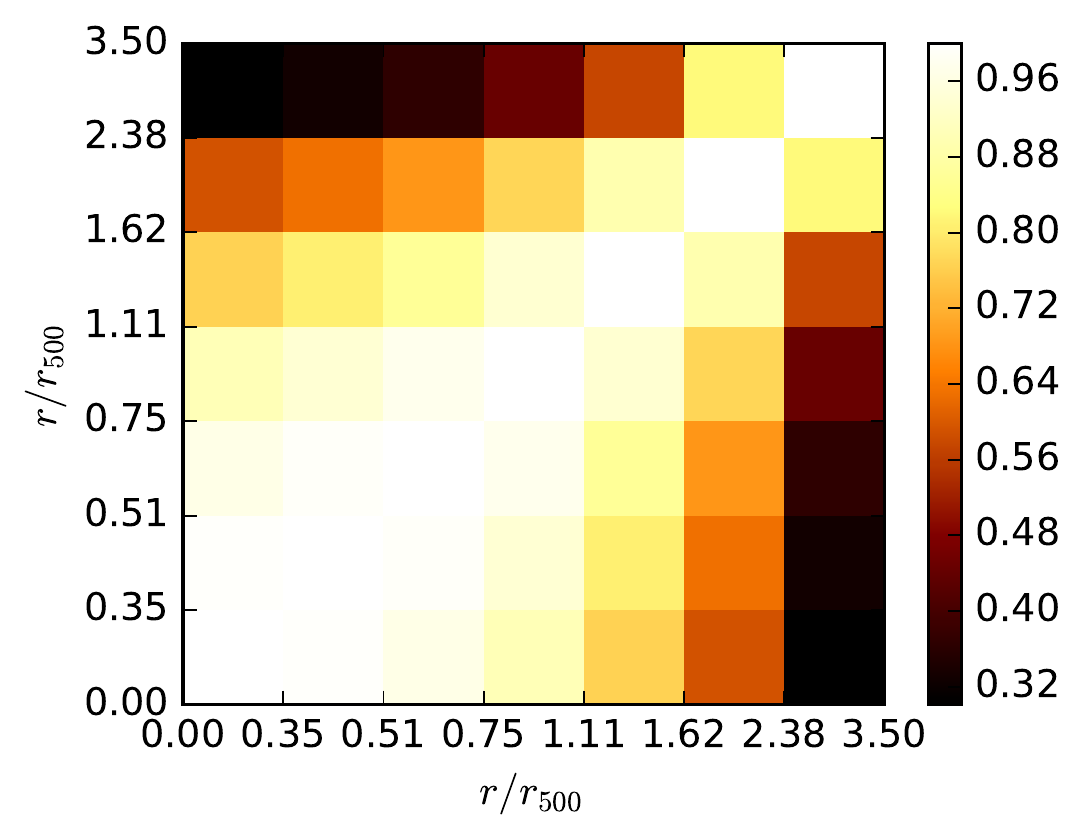}}
\caption{
Normalized covariance matrix used in the likelihood analysis for a mass of $M_{500} = 5\times10^{14} M_\odot$, corresponding to $r_{500}=1.11\,\mathrm{Mpc}$. At the mean redshift of the sample and for the same mass, $r/r_{500}=3.5$ corresponds to 13.5 $\mathrm{arcmin}$. The 30\% base level correlation is coming from the fact that in each map the same region is used to subtract the background from all the radial bins.
}
\label{fig:profile_covariance_matrix}
\end{figure}

The covariance matrixes $\mathcal{C}$ are estimated from $N_R=$ 1,000 patches randomly chosen within the redMaPPer footprint, where the same measurement described above is done. As this measurement depends on the $M_{500}$ value proposed, the $7\times7$ covariance matrix is recomputed each time as:
\begin{equation}
\mathcal{C}_{ij} (M_{500})=\frac{1}{N_R-1}\sum_{n=1}^{N_R}(y_i^n - \left<y_i\right>)(y_j^n - \left<y_j\right>)\,.
\end{equation}
As an example, the covariance matrix obtained considering a mass of $M_{500} = 5\times10^{14} M_\odot$ is shown in Fig.~\ref{fig:profile_covariance_matrix}.

%================================================
%================================================

Finally, we consider the gas fraction constraints introducing a Gaussian prior $\sim\mathcal{N}(f_{\rm gas},\, \sigma_{f_{\rm gas}})$, where $f_{\rm gas}$ is estimated from the results of \cite{Pratt2009}, as explained above. In each MCMC step we compute, following the procedure described in Sec.~\ref{sect:gas_fraction}, the 6 gas fractions associated to a given set $\bm{\phi}$ of pressure profile parameters and masses, and then use Eq.~\ref{eq:pratt_gas_fraction} to model the expected gas fraction $f_{\rm gas}$ for each value of $M_{500}^k$, which we use for the prior. To estimate $\sigma_{f_{\rm gas}}$, we add in quadrature the errors derived from the uncertainties on both the mass--gas fraction (Eq.~\ref{eq:pratt_gas_fraction}) and the mass--temperature (Eq.~\ref{eq:MT_relation}) relations, which are obtained propagating through a Monte Carlo (MC) method. Because we have decided to be as conservative as possible on the relations employed and the resulting uncertainties in this analysis are large, we notice that the contribution that this prior has in the final estimated values of the pressure profile is small, only limiting those models where the gas fraction $f_{{\rm gas},500}$ takes values below 0.05 or above 0.20 for masses in the $\sim 10^{14}-10^{15}\,M_\odot$ range.

For all $P_0$, $c_{500}$, $\alpha$, $\beta$, $M_{500}^1,\,...\,,\,M_{500}^{6}$, we consider flat uninformative priors $\sim\mathcal{U}(\infty,\,-\infty)$, allowing for a wide range of different model-masses configurations.

\subsection{Results}

\begin{figure*}
\centering
\includegraphics[width=0.98\textwidth]{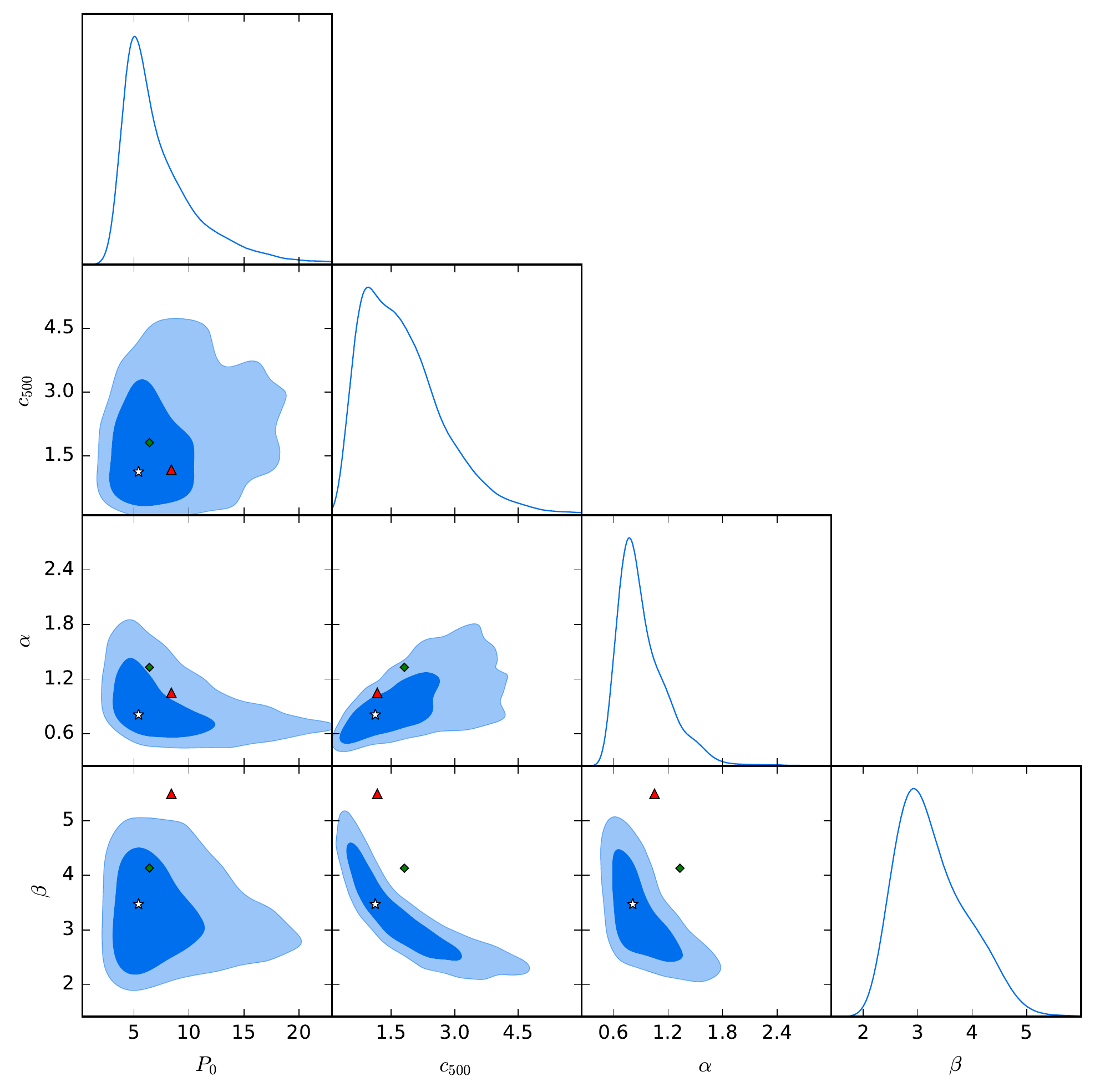}
\caption{
Marginalised posterior distributions of the universal pressure profile parameters $P_0$, $c_{500}$, $\alpha$ and $\beta$, as obtained in our MCMC analysis. Contours represent 68\% and 95\% confidence levels. The best fit values, corresponding to $[P_0,\,c_{500},\,\gamma,\,\alpha,\,\beta]=\UPPbf$, are marked with a white star. The best fit obtained by Arnaud et al. (2010), [8.40 $h_{70}^{-3/2}$, 1.18, 0.31, 1.05, 5.49], is marked with a red triangle, and Planck Collaboration et al. (2013) best fit, [6.41, 1.81, 0.31, 1.33, 4.13], with a green diamond.
}
\label{fig:likelihood_profile_2D}
\end{figure*}

The derived posterior probabilities of the GNFW universal pressure profile parameters $P_0$, $c_{500}$, $\alpha$ and $\beta$ are displayed in Fig.~\ref{fig:likelihood_profile_2D}. To compute the center (mean) and the scale (dispersion) of the marginalised posterior distributions, we use the robust estimators described in \cite{Beers1990}. The values obtained with this method, together with the best fit values, are listed in Table~\ref{tab:likelihood_profile}.

%================================================
%================================================

%================================================
\begin{table}
\center
\caption{
\label{tab:likelihood_profile}
The recovered values of the GNFW universal pressure profile parameters, together with the best fit values.
}
\begin{tabular}{c c c}
\hline
{\bf Parameter}	&{\bf Mean value} 	&{\bf Best fit}\\
\hline
\hline
$P_0$ 		&$6.48 \pm 3.00$	&$5.42$\\
$c_{500}$ 	&$1.65 \pm 0.95$	&$1.12$\\
$\alpha$ 	&$0.87 \pm 0.25$	&$0.81$\\
$\beta$ 	&$3.19 \pm 0.66$	&$3.47$\\
\hline
\end{tabular}
\end{table}
%================================================

\begin{figure*}
\centering
\includegraphics[width=0.98\textwidth]{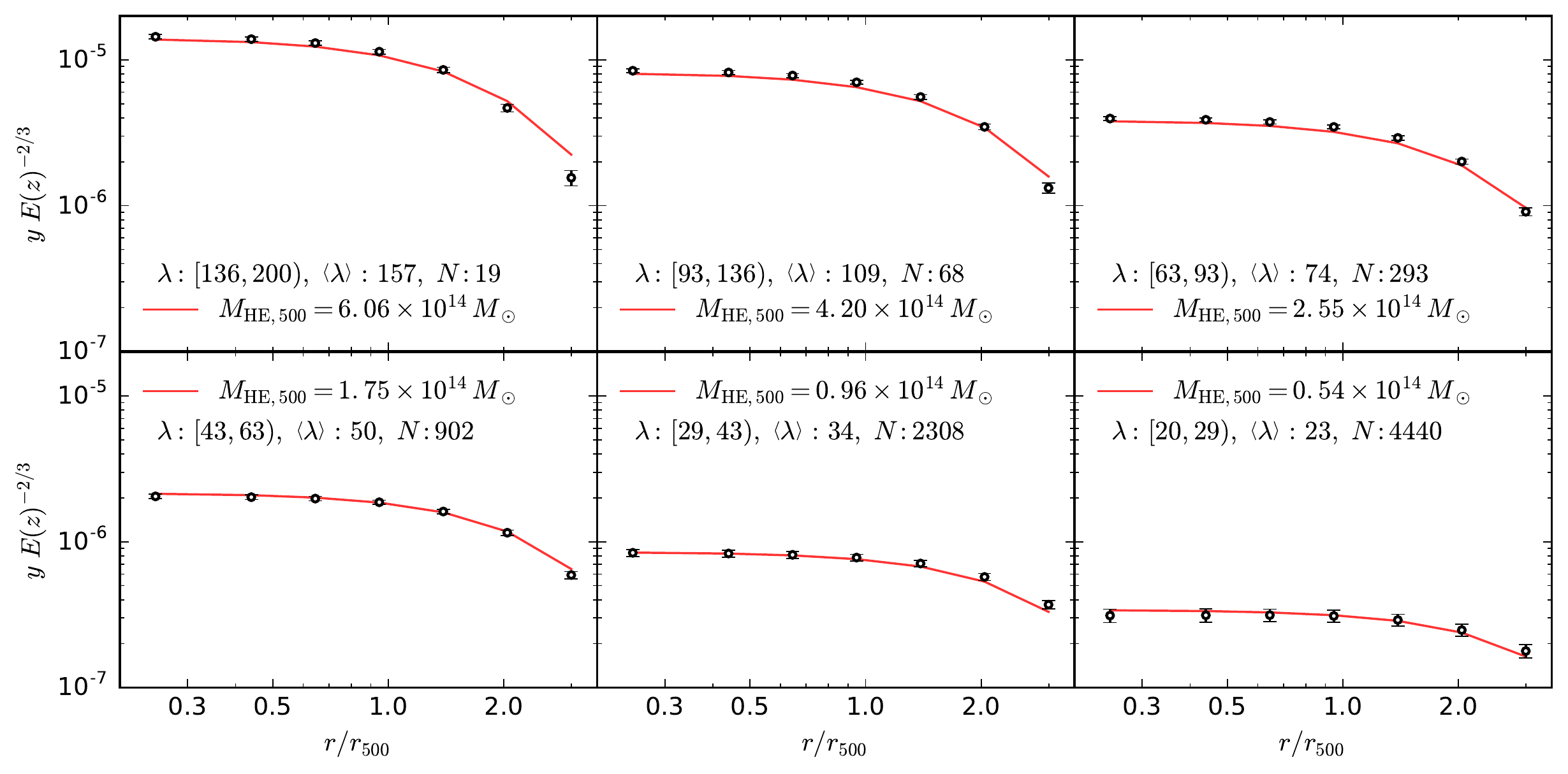}
\caption{
The values of $y$ recovered at different scaled radii for the cluster subsets divided by richness, and ordered as in Fig.~\ref{fig:y_maps}. The error bars are the square root of the diagonal elements of the covariance matrix. The red lines represent the prediction from the universal pressure profile with the best fit parameters, $[P_0,\,c_{500},\,\gamma,\,\alpha,\,\beta]=\UPPbf$. The best fit masses used for the model are also shown for each sample, together with the richness range, the mean richness, and the number of clusters contained in them. It should be noted that, as we are not deconvolving the observed profiles in our analysis, they are not necessarily self-similar due to the different relative size of the {\it Planck} beam with respect to $r_{500}$ for the 6 cluster subsamples.
}
\label{fig:y_profiles}
\end{figure*}

The $y$ radial profiles recovered for the cluster subsamples considered are displayed in Fig.~\ref{fig:y_profiles}, together with the joint best fit $y$ profile model obtained, as shown in Table~\ref{tab:likelihood_profile}.

%================================================
%================================================

Finally, the mean masses recovered for the cluster subsamples are listed in Table~\ref{tab:likelihood_masses}.

%================================================
\begin{table}
\center
\caption{
\label{tab:likelihood_masses}
Richness range, number of clusters, mean richness and mean masses recovered from the SZ signal of the cluster subsamples studied.
}
\begin{tabular}{c c c c}
\hline
{\bf Richness range}	&\bm{$N$}&\bm{$\left<\lambda\right>}$	&\bm{ $\left<M_{{\rm HE,}500}\right>$ $[10^{14}\,M_\odot]$}\\
\hline
\hline
$[136,\,200)$ 		&19 	&157.6			&$6.48 \pm 0.66$\\
$[93,\,136)$ 		&68 	&109.6			&$4.52 \pm 0.42$\\
$[63,\,93)$ 		&293 	&74.3			&$2.77 \pm 0.25$\\
$[43,\,63)$ 		&902 	&50.8			&$1.84 \pm 0.18$\\
$[29,\,43)$ 		&2308 	&34.6			&$1.04 \pm 0.13$\\
$[20,\,29)$ 		&4440 	&23.8			&$0.54 \pm 0.12$\\
\hline
\end{tabular}
\end{table}
%================================================

\subsubsection{Comparison with previous results}

For this model ($[P_0,\,c_{500},\,\gamma,\,\alpha,\,\beta]$), and using the {\it XMM-Newton} X-ray data up to $r/r_{500} \lesssim 1$ of a sample of 33 X-ray selected local $(z<0.2)$ clusters covering the $10^{14}M_\odot < M_{500} < 10^{15}M_\odot$ mass range, \cite{Arnaud2010} obtained the best fit values [8.403 $h_{70}^{-3/2}$, 1.177, 0.3081, 1.0510, 5.4905], relying on numerical simulations for larger radii.

Deconvolving the stacked pressure profile of 62 SZ selected clusters, the \cite{Planck2013V} obtained, combining {\it Planck} with {\it XMM-Newton} data, the best fit values [6.41, 1.81, 0.31, 1.33, 4.13], where the \cite{Arnaud2010} value of $\gamma=0.31$ had been previously fixed, as we have done in our analysis. By comparison with pure estimates, it can be seen in Fig.~5 of \cite{Planck2013V}, that a large uncertainty is present in the estimation of these parameters, with a high degeneracy between them, which we attribute to the larger radial extent of our SZ profiles and the wider cluster mass coverage of our sample.

We also notice that the external slope, $\beta$, derived in this work points to shallower profiles in the outer part of the clusters. This is in agreement with the results derived in the Coma and Virgo clusters \citep{PlanckX2013, Planck2016XL} based on \textit{Planck} data where the SZ signal extends to beyond the virial radius in those clusters. Like in those papers, we can reach similar distances from the virial radii and be sensitive to the external slope of clusters where the signal from neighbouring merging filaments is expected to flatten the SZ profile.

\section{Mass bias and mass--richness relation}
\label{sect:mass--richness_relation_analysis}

\subsection{Likelihood analysis}

We now obtain the value of the mass bias comparing our results with recent weak lensing mass derivations that make use of stacked subsamples of the redMaPPer cluster catalogue. We consider the results by \citeauthor{Simet2016} (\citeyear{Simet2016}, S16 from now on), who, making use of SDSS data for the weak lensing mass estimates of the redMaPPer clusters in the $0.10 < z < 0.33$ redshift region, obtained one of the most precise mass--richness relations to date, and those by \citeauthor{Melchior2016} (\citeyear{Melchior2016}, M16 hereafter), which, using DES SV data to make stacked measurements of the weak lensing shear as a function of mean cluster richness and mean cluster redshift, measured a redshift-dependent mass--richness relation of comparable precision to that of S16. Because both of them are given in terms of $M_{200m}$, we consider a NFW profile \cite{Navarro1996} and the mass--concentration relation of \cite{Bhattacharya2013} to convert between different mass definitions.

At the same time, we derive the optimal mass--richness relation able to describe our data, considering the probability distribution of the bias obtained from the comparison with S16 and M16 results. To do so, we first compare our masses with the masses estimated by S16 and M16 with a joint likelihood:
\begin{equation}
\ln\mathcal{L}_{\rm bias} = \ln\mathcal{L}_{\rm S16} + \ln\mathcal{L}_{\rm M16}\,,
\end{equation}
where $\mathcal{L}_{\rm S16}$ and $\mathcal{L}_{\rm M16}$ are computed comparing our 6 masses with the masses predicted by the S16 and the M16 mass--richness relations, with $\ln\mathcal{L} \propto -\chi^2/2$, and:
\begin{equation}\label{eq:chi_square_MR}
\chi^2\left[(1-b)\right] = \left(\bm{M_{500}} - \bm{M^\text{\bf\tiny model}_{500}}\right) \mathcal{C}^{-1} \left(\bm{M_{500}} - \bm{M^\text{\bf\tiny model}_{500}}\right)^T\,,
\end{equation}
where the bias-corrected masses are computed as $M_{500} = M_{{\rm HE,}500}/(1-b)\,$, following Eq.~\ref{eq:bias_factor}, and the S16 and M16 $M^\text{\tiny model}_{500}$ model masses are obtained evaluating the corresponding S16 or M16 mass--richness relation at the 6 cluster subsamples mean richnesses.

To obtain the resulting bias-dependent mass--richness relation, we compare these bias-corrected masses with the masses predicted by a generic mass--richness relation, which we compute following the steps described in Sec.~\ref{sect:mass--richness_relation_model} and considering a given set of free parameters $\log_{10} M_0$, $\alpha_{M|\lambda}$ and $\sigma_{M|\lambda}$, which we constrain. The global likelihood has the form:
\begin{equation}
\ln\mathcal{L} = \ln\mathcal{L}_{\rm bias} + \ln\mathcal{L}_{M|\lambda}\,,
\end{equation}
where $\mathcal{L}_{M|\lambda}$ is computed now through another chi-square function similar to that of Eq.~\ref{eq:chi_square_MR}, but considering both the value of the bias and the values of the mass--richness relation parameters for the model masses.

Because the values of the SZ-estimated masses are correlated, this new covariance matrix $\mathcal{C}$ is obtained directly from the MCMC analysis performed in Sec.~\ref{sect:sz--mass_relation_analysis}, and used in this new MCMC likelihood calculation. We also include the errors coming from the S16 and M16 model uncertainties, adding them in quadrature to the diagonal of the covariance matrix.

For $\log_{10} M_0$, $\alpha_{M|\lambda}$ and $(1-b)$ we assume flat uninformative priors, meanwhile for $\sigma_{M|\lambda}$ we assume the inverse gamma distribution prior $\sim IG(\epsilon,\,\epsilon)$, with $\epsilon=10^{-3}$ \citep{Andreon2010}.

\subsection{Results}

We notice that this analysis, based on mean masses rather than individual measurements, is not able to constrain the value of $\sigma_{M|\lambda}$. The value of the scatter, difficult to constrain in general, is usually found to be between 0.15 and 0.30. \cite{Saro2015} find $\sigma_{\ln M|\lambda^{\rm obs}}=0.18^{+0.08}_{-0.05}$, meanwhile \cite{Rozo2014a} and \cite{Rozo2015a}, comparing individual redMaPPer clusters with X-ray and SZ mass estimates, find a value $\sigma_{\ln M|\lambda^{\rm obs}} \approx 0.25\pm0.05$. 

\begin{figure}
\resizebox{84mm}{!}{\includegraphics{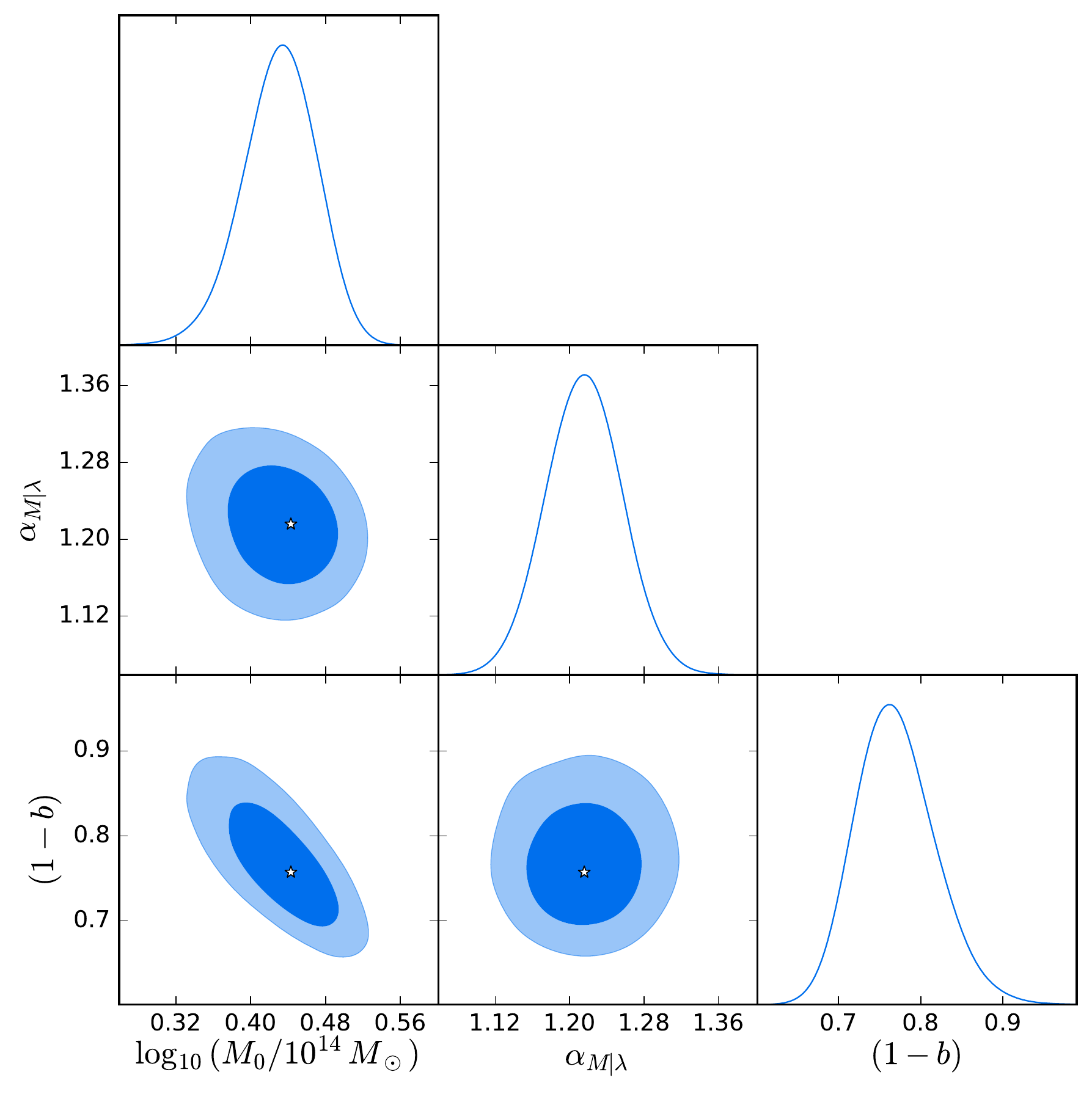}}
\caption{
Posterior probability distributions of the mass--richness parameters $\log_{10} M_0$ and $\alpha_{M|\lambda}$, and of the bias factor $(1-b)$. The best fit location is marked with a white star. The mass--richness relation scatter, $\sigma_{M|\lambda}$, cannot be constrained by our data and its posterior probability distribution is greatly influenced by the posterior, and thus not shown.
}
\label{fig:likelihood_MR_2D}
\end{figure}

In any case, the posterior probability of $\log_{10} M_0$, $\alpha_{M|\lambda}$ and $(1-b)$ recovered from the MCMC is shown in Fig.~\ref{fig:likelihood_MR_2D}. As in the previous section, we use the \cite{Beers1990} estimator to obtain the values from the marginalised posterior distributions. The results are shown in Table~\ref{tab:likelihood_MR}.

%================================================
%================================================

%================================================
\begin{table}
\label{tab:likelihood_MR}
\center
\caption{
The confidence values of the mass--richness relation parameters and of the bias factor $(1-b)$, together with the best fit. $\alpha_{M|\lambda}$ is the slope of the mass--richness relation, and $\log_{10}M_0$ is the pivot mass, evaluated at $\lambda_0=60$ and given in terms of $M_{500c}$.
%For the $\sigma_{M|\lambda}$ parameter $IG$ prior, a value of $\epsilon=10^{-3}$ has been taken. 
The stacking analysis performed is not able to constrain the value of $\sigma_{M|\lambda}$, which is largely determined by the prior, and thus not shown.
}
\begin{tabular}{c c c}
\hline
{\bf Parameter}			&{\bf Mean value}	&{\bf Best fit}\\
\hline
\hline
$\log_{10}\,(M_0/M_\odot)$	&$14.435 \pm 0.040$	&$14.446$\\
$\alpha_{M|\lambda}$		&$1.22 \pm 0.04$	&$1.21$\\
$(1-b)$				&$0.76 \pm 0.05$	&$0.75$\\
$\sigma_{M|\lambda}$		&-			&-\\
\hline
\end{tabular}
\end{table}
%================================================

\subsubsection{Comparison with previous results}

We now compare the value of the bias that we recover with those found in the literature. We also compare the associated bias-dependent mass--richness relation with those mass--richness relations used to obtain it, and with the one obtained recently by \cite{Baxter2016}. We would like to remind the reader that our estimations of both the bias and the mass--richness relation depend on the results of S16 and M16, and should not be considered independent of them.

Comparing ``observed'' with ``true'' masses coming from several numerical simulations, \cite{Planck2013XX} derived a mass-dependent bias with a mean value of $(1-b)=0.8^{+0.2}_{-0.1}$. However, a bias of the order of 40 per cent was needed in order to reconcile {\it Planck} clusters counts with CMB observations, and its value is still the focus of intense research. \cite{vonderLinden2014}, comparing {\it Planck} cluster mass estimates with weak lensing masses from the Weighing the Giants (WtG) project, observed a higher bias of $(1-b)=0.70\pm0.06$. In a similar way but using 50 clusters from the Canadian Cluster Comparison Project (CCCP), \cite{Hoekstra2015} found a value of $(1-b)=0.76\pm0.05\pm0.06$. On the other hand, \cite{Smith2016} obtained a smaller bias of the order of the 5 per cent from a sample of 50 clusters with X-ray and weak lensing masses in the $0.15 < z < 0.3$ redshift range, being statistical uncertainties more relevant than any bias, and claimed that the high values of the bias inferred by the WtG and CCCP samples may be dominated by clusters at $z>0.3$. Although \cite{Sereno2016} found a value of the bias of the order of 25 per cent for {\it Planck} masses with respect to their weak lensing masses, it was claimed that this value was strongly dependent on redshift, following \cite{Smith2016} results. \cite{Saro2016}, following an approach similar to ours but using stacked measurements of the SZ signal of 719 DES redMaPPer clusters with South Pole Telescope (SPT) data, needed a bias as high as $(1-b)=0.52\pm0.05$ to follow the model predictions when the \cite{Arnaud2010} pressure profile was assumed. However, this bias was reduced to the range $(1-b)=0.7 - 0.9$ when only the clusters with richness $\lambda>80$ were taken into account, or when other scaling relations were considered. More recently, and using a sample of 35 {\it Planck} clusters that were within the area covered by the CFHTLenS and RCSLenS photometric surveys, \cite{Sereno2017} found that the {\it Planck} estimated masses were biased low by $\sim27\pm11$ per cent with respect to weak lensing masses, consistent with our findings here.

%Melchior2016
Regarding the mass--richness relation, in M16 a redshift-dependent mass--richness relation was derived with a slope of $1.12\pm0.20\pm0.06$ for a pivot mass $\log_{10}\left(M_{200m}/M_\odot\right)=14.371 \pm 0.040 \pm 0.022$ at a pivot richness $\lambda=30$ and $z=0.5$. Evaluated at our mean redshift $z=0.245$ and their pivot richness $\lambda=30$, their mass--richness relation gives a mass $\log_{10}\left(M_{200m}/M_\odot\right)= 14.364 \pm 0.080$. Converting our masses to $M_{200m}$, we find at $\lambda=30$ a value of $\log_{10}\left(M_{200m}/M_\odot\right)= \logMMelchior $.
%Simet2016
In the S16 case, they obtained a mass--richness relation with a slope equal to $1.33^{+0.09}_{-0.10}$ for a pivot mass of $\log_{10}(M_{200m}/h^{-1}M_\odot)=14.344 \pm 0.021 \pm 0.023$ at $\lambda=40$. Converting to their units, we find that our mass--richness relation at their pivot richness yields a mass equal to $\log_{10}(M_{200m}/h^{-1}M_\odot)= \logMSimet $. 
%Baxter2016
\cite{Baxter2016}, using the angular clustering of redMaPPer clusters confined in two redshift bins in the $0.18 < z < 0.33$ redshift region, found a value of the slope of the mass--richness relation of $1.18 \pm 0.16$ and a pivot mass of $\ln(M_{200m}/M_\odot)=33.66\pm0.18$ at $\lambda=35$ and $z=0.25$. In our mass--richness relation at their pivot richness, we find $\ln(M_{200m}/M_\odot)= \logMBaxter$.

\begin{figure}
\resizebox{84mm}{!}{\includegraphics{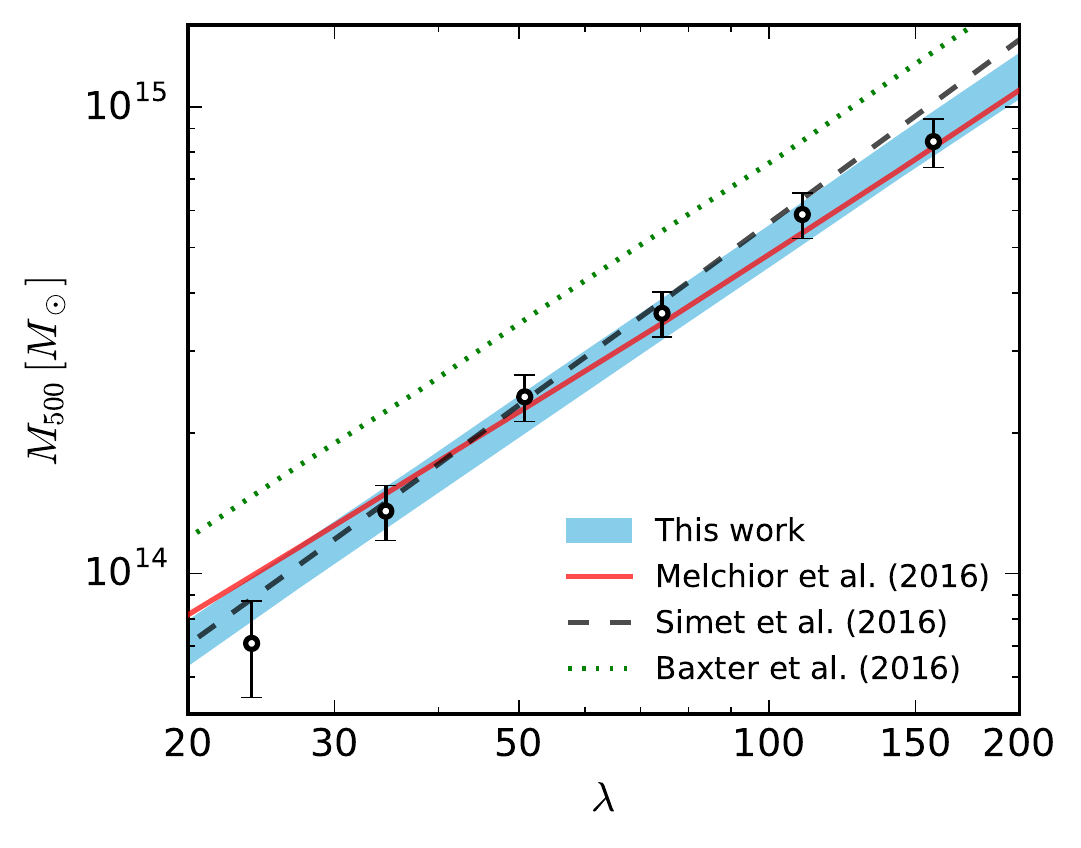}}
\caption{
The blue shaded region represents our $1\sigma$ confidence interval of the mean mass--richness relation with the parametrisation shown in Table~\ref{tab:likelihood_MR}, derived from the mean bias-corrected masses (black circles) of the cluster subsamples considered, displayed as a function of the original sample mean richness. The errors on the $M_{500}$ masses include the uncertainty on the value of the bias. 
The two mass--richness relations that we have used to obtain an estimate of the bias, namely those derived by Melchior et al. (2016, M16) (red solid line) and Simet et al. (2016, S16) (black dashed line), are shown, together with Baxter et al. (2016) mass--richness relation (green dotted line).
}
\label{fig:mass_vs_richness}
\end{figure}

Our derived mass--richness relation, together with the bias-corrected masses used in this analysis and the redMaPPer-based mass--richness relations found in the literature and just mentioned, are shown in Fig.~\ref{fig:mass_vs_richness}. Our estimated mass-richness relation can be seen to bracket the work of \cite{Simet2016} and \cite{Melchior2016}, in good agreement. Whereas the \cite{Baxter2016} agrees in terms of the slope of the relation, it lies significantly above the other work, including our own.

%================================================
%================================================

\section{$Y_{500}$--$M_{500}$ relation}
\label{sect:y500_m500_relation}

We now straightforwardly derive the $Y_{500}$--$M_{500}$ relation by considering the values of $P_0$, $c_{500}$, $\alpha$, $\beta$ and $(1-b)$ obtained, and assuming again a value of $\gamma=0.31$. Following the usual notation we have:
\begin{equation}
E^{-2/3}(z)\left[\frac{Y_{500}}{10^{-4}\,{\rm Mpc}^2}\right] = 10^A\left[\frac{(1-b)\,M_{500}}{6\times 10^{14}M_\odot}\right]^B\,,
\end{equation}
in which the redshift evolution is considered to be self-similar \citep{Kaiser1986}. From the results previously obtained, we find:
\begin{equation}\nonumber
\begin{aligned}
A &= \AYM\,,\\
B &= \BYM\,.
\end{aligned}
\end{equation}
The value of the slope is actually in good agreement with the expectation from self-similarity, $B\sim5/3$. This value of the slope is within $1\sigma$ from the results of $B = 1.79 \pm 0.08$ obtained in \cite{Planck2013XX}, with a normalisation factor of $A=0.19\pm0.02$, different from our value of $A$ but compensated by the difference in the $(1-b)$ factor. Our result is also compatible with those of \cite{Sereno2015a}, who obtained values for the slope of 1.4 - 1.9, albeit a relatively wide range of slope, by considering samples of clusters with weak lensing mass estimates.

It should be noted that this is the mean $Y_{500}$--$M_{500}$ relation, so, although small in the SZ case, a scatter $\sigma_{Y|M}$ should be considered when applying it to individual clusters, just as when converting observed richness into mass (Eqs.~\ref{eq:MR_relation_2} and \ref{eq:MR_relation_3}). In \cite{Planck2013XX} a value of of the scatter of $\sim15$ per cent is considered, meanwhile in \cite{Sereno2015a} a scatter of the the 15 - 30 per cent order is estimated.

Finally, as a consistency check on our results, we measure $Y^{\rm cyl}(3.5\,r_{500})$ in the 6 $y$ maps corresponding to the 6 cluster subsamples, deriving the values of $r_{500}$ from the mean masses obtained in the MCMC analysis, listed in Table~\ref{tab:likelihood_masses}, and converting to $Y_{500}$ using the universal pressure profile as given by the $P_0$, $c_{500}$, $\alpha$ and $\beta$ parametrisation of Table~\ref{tab:likelihood_profile}. We derive the errors from 1,000 random patches within the redMaPPer footprint in a similar way as we did in Sec.~\ref{sect:likelihood_analysis_1}. We propagate the errors coming from the uncertainty in the pressure profile parameters, the masses and the bias factor drawing a large enough number of samples from the posterior probability distribution of the $4+6+1$ dimensional parameter space, repeating the measurements, and adding in quadrature the variance of the $Y^{\rm cyl}(3.5\,r_{500})$ values obtained.

\begin{figure}
\resizebox{84mm}{!}{\includegraphics{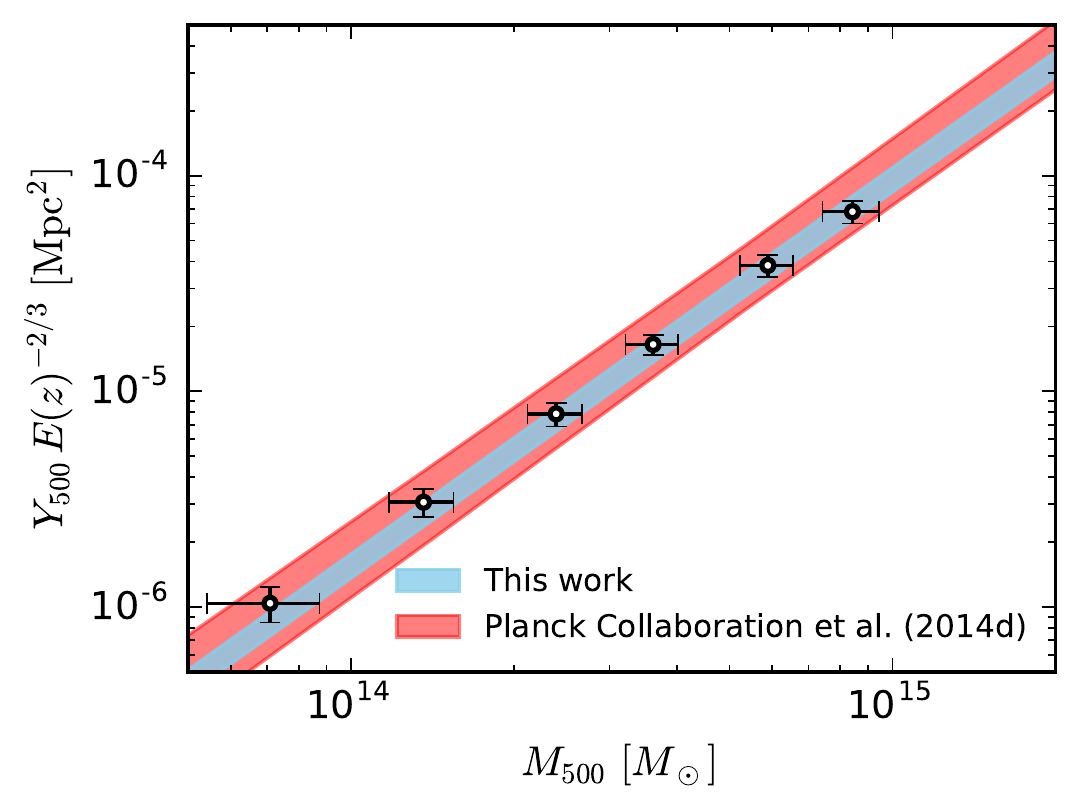}}
\caption{
The blue shaded region shows our 1$\sigma$ confidence interval of the scaling relation between $Y_{500}$ and $M_{500}$, as given by the posterior probability distribution of the pressure profile parameters $P_0$, $c_{500}$, $\alpha$ and $\beta$, the masses, and the bias factor. The black circles dots are the observed $Y^{\rm cyl}(3.5\,r_{500})$ values for the cluster subsamples, converted to $Y_{500}$ using the pressure profile parameters of Table~\ref{tab:likelihood_profile} and the masses of Table~\ref{tab:likelihood_masses}. The errors on the $M_{500}$ masses include the uncertainty coming from the bias. As the red shaded region, we plot the Planck Collaboration et al. (2014d) $Y_{500}$--$M_{500}$ relation for comparison, where we have considered not only the errors in the slope and the normalisation, but also the uncertainty coming from the bias factor correction.
}
\label{fig:Y500_vs_M500}
\end{figure}

Our mean $Y_{500}$--$M_{500}$ relation and the measured values of $Y_{500}$ are shown in Fig.~\ref{fig:Y500_vs_M500}, where we also plot the relation considered in \citet{Planck2013XX}, with which good agreement is found. In comparison our work follows a tighter relation, which is attributed to the relatively large size of our cluster sample.

%================================================
%================================================

%%%%%%%%%%%%%%%%%%%%%%%%%%%%%%%%%%%%%%%%%%%%%%%%%%%%%%%%%%%%%%%%%%%%%%%%%%%%%%%
\section{Summary and conclusions}
\label{sect:conclusions}

In this chapter we have presented and employed a new method to constrain simultaneously the GNFW universal pressure profile parameters and the masses of 6 different richness subsamples of clusters using stacked measurements of the SZ effect. We then estimated both a mass--richness and a $Y_{500}$--$M_{500}$ relation using weak lensing mass estimates found in the literature.

Using the positions in the sky of $\sim$ 8,000 redMaPPer clusters in the $0.100 < z < 0.325$ volume-complete redshift region, we have produced and stacked the {\it Planck} full mission SZ maps of 6 richness subsamples of clusters, and constrained the common GNFW universal pressure profile parameters and the mean masses of each subsample through a MCMC analysis, obtaining better constraints than previous works, with $P_0 = \Pval$, $c_{500} = \cval$, $\alpha = \alphaval$ and $\beta = \betaval$ for a fixed value of $\gamma=0.31$, and masses in the $0.7\times10^{14}\,M_\odot \lesssim M_{500} \lesssim 9\times10^{14}\,M_\odot$ range. The universal pressure profile parameters best fit is found at $[P_0,\,c_{500},\,\gamma,\,\alpha,\,\beta]=\UPPbf$. To improve the precision of our analysis, we have used as a prior the \cite{Pratt2009} mass--gas fraction relation obtained from different subsets of {\it Chandra} and {\it XMM-Newton} clusters.

We also notice that the external slope, $\beta$, derived in this work points to shallower profiles in the outer part of the clusters. This is in agreement with the results derived in the Coma and Virgo clusters \citep{PlanckX2013, Planck2016XL} based on \textit{Planck} data where the SZ signal extends to beyond the virial radius in those clusters. Like in those papers, we can reach similar distances from the virial radii and be sensitive to the external slope of clusters where the signal from neighbouring merging filaments is expected to flatten the SZ profile.

Then we have compared our SZ-based masses with independent weak lensing mass estimates from the mass--richness relations of \citeauthor{Simet2016} (\citeyear{Simet2016}, S16) and \citeauthor{Melchior2016} (\citeyear{Melchior2016}, M16), finding a mean mass bias of $(1-b)=\bias$. This value does not depend strongly on the richness subsample considered, and is in good agreement with other independent recent estimates of the HE bias. Because we are working in the $0.100<z<0.325$ redshift region it may not be surprising that we differ with the higher redshift base estimate of \cite{Smith2016}, but this may be understood given their claim that this bias may be enhanced in the $z>0.3$ region.

By correcting for our estimated HE bias, we have then straightforwardly derived the mass--richness relation able to describe the bias-corrected masses, finding a slope equal to $\alpha_{M|\lambda} = \alphaMR$ for a pivot mass $\log_{10}\,(M_{500c}/M_\odot) = \logMRval$ evaluated at $\lambda=60$ at the mean redshift of the sample, $z=0.245$. The amplitude of the mass--richness relation is strongly tied to the weak lensing-based HE bias correction results by S16 and M16 as described above, whereas the slope of the mass-richness relation that we find is not linked because the bias correction is found to be independent of cluster mass over our observed range.

From the estimated pressure profile parameters and bias, we have then derived a mean $Y_{500}$--$M_{500}$ relation $E^{-2/3}(z)\,\left<Y_{500}\right> \propto 10^A\,\left<(1-b) M_{500}\right>^B$ with a normalisation factor $A =\AYM$ and a slope $B = \BYM$. When the contribution from the bias is considered, this result is within 1$\sigma$ from the \textit{Planck} mean relation in all the mass range considered, but has a tighter range because of the relatively large size of our sample.

The results obtained in this work show the potential of this method, which provides powerful tools to improve the estimation of cosmological parameters using clusters, which will be crucial in future and ongoing surveys like the J-PAS \citep{Benitez2014} or DES \citep{DES2016} surveys.

%%%%%%%%%%%%%%%%%%%%%%%%%%%%%%%%%%%%%%%%%%%%%%%%%%%%%%%%%%%%%%%%%%%%%%%%%%%%%%%
\hspace{0pt}\\ \textbf{Acknowledgements}\\

PJ would like to thank the Astrophysics Group of UCL, specially Ofer Lahav, for the opportunity to visit and learn from them.
IDM acknowledge financial supports from University of the Basque Country UPV/EHU under the program ``Convocatoria de contrataci\'{o}n para la especializaci\'{o}n de personal 
investigador doctor en la UPV/EHU 2015",
and from  the Basque Government through the research project IT-956-16.
TJB is supported by IKERBASQUE, the Basque Foundation for Science.
J.M.D acknowledges support of the projects AYA2015-64508-P (MINECO/FEDER, UE), AYA2012-39475-C02-01 and the consolider project CSD2010-00064 funded by the Ministerio de Economia y Competitividad.
RL and IDM are supported by the Spanish Ministry of Economy and Competitiveness through research projects FIS2010-15492.
RL is also supported by Consolider EPI CSD2010-00064, and the University of the Basque Country UPV/EHU under program UFI 11/55.
PJ acknowledges financial support from the Basque Government grant BFI-2012-349.
TJB, RL and PJ are also supported by the Basque Government through research project GIC12/66.

Funding for SDSS-III has been provided by the Alfred P. Sloan Foundation, the Participating Institutions, the National Science Foundation, and the U.S. Department of Energy Office of Science. The SDSS-III web site is http://www.sdss3.org/. SDSS-III is managed by the Astrophysical Research Consortium for the Participating Institutions of the SDSS-III Collaboration including the
University of Arizona,
the Brazilian Participation Group,
Brookhaven National Laboratory,
University of Cambridge,
Carnegie Mellon University,
University of Florida,
the French Participation Group,
the German Participation Group,
Harvard University,
the Instituto de Astrofisica de Canarias,
the Michigan State/Notre Dame/JINA Participation Group,
Johns Hopkins University,
Lawrence Berkeley National Laboratory,
Max Planck Institute for Astrophysics,
Max Planck Institute for Extraterrestrial Physics,
New Mexico State University,
New York University,
Ohio State University,
Pennsylvania State University,
University of Portsmouth,
Princeton University,
the Spanish Participation Group,
University of Tokyo,
University of Utah,
Vanderbilt University,
University of Virginia,
University of Washington,
and Yale University.

\bibliography{BiblioRef}{}

\label{lastpage}

\end{document}